\documentclass[aps,prl,twocolumn,floatfix,showpacs,10pt]{revtex4-1}

\usepackage{latexsym}
\usepackage{graphicx}
\usepackage{times,psfrag}
\usepackage{amsmath}
\usepackage{dcolumn}
\usepackage{color}
\usepackage{amssymb,bm,euscript}
\usepackage{amsfonts,hyperref}
\usepackage{wasysym}
\usepackage{subfigure}

\newcommand{\bs}{\boldsymbol}

\begin{document}
\title{Frustrated quantum critical theory of putative spin-liquid phenomenology in 6H-B-Ba$_3$NiSb$_2$O$_9$}

\author{G. Chen}
\affiliation{Department of Physics, University of Colorado, Boulder, CO 80309, USA}
\author{M. Hermele}
\affiliation{Department of Physics, University of Colorado, Boulder, CO 80309, USA}
\author{L. Radzihovsky}
\affiliation{Department of Physics, University of Colorado, Boulder, CO 80309, USA}

\begin{abstract}
A recently discovered material, 6H-B-Ba$_3$NiSb$_2$O$_9$ was found to display unusual 
low-temperature phenomenology, interpreted as a quantum spin liquid with spin $S=1$
on a triangular lattice. We study a spin $S=1$ exchange model on an AB stacked triangular lattice near its 
quantum paramagnet-to-spiral transition, driven by easy-plane single-ion anisotropy.  We demonstrate 
that the frustrated inter- and intra-layer exchanges induce contour lines of low-energy excitations that lead to a broad crossover regime of linear-temperature dependence of the specific heat. Based
on this and various other predictions, we argue that the observed phenomenology can be understood in
terms of a conventional picture of a proximity to this frustrated critical point.
\end{abstract}
\date{\today}


\maketitle

Quantum spin liquids (QSLs) are Mott insulators that remain magnetically disordered down to zero temperature, and, as we use term here, are exotic states of matter characterized by properties such as quantum number fractionalization, topological order, and gapless excitations in the absence of spontaneously broken symmetry.  The realization of  QSLs in theoretical models has been well established\cite{Balents10}, and a number of materials have emerged as promising candidates\cite{Helton07,Shimizu03,Itou08,Okamoto07,Okamoto09,Hiroi01,Vries10,Gardner99}. However, there is no direct confirmation of QSL in any of these systems, and alternative explanations now exist for some QSL candidates\cite{Kohno07,Starykh10,Stoudenmire09,Chen09,Singh10}.

Many QSL candidates share a rough phenomenology: they are electrical insulators, but with thermodynamic properties similar to those of a metal.  In particular, many of these systems have a constant low-temperature spin susceptibility, and a linear-temperature dependence of the low-temperature specific heat.  Theoretical attempts to explain this behavior usually invoke spin-$\frac{1}{2}$ fermionic spinons with a constant density of states (DOS). In this Letter, we propose the first (to our knowledge) alternative explanation for this phenomenology that does not invoke substantial quenched disorder.

Recently the compound 6H-B-Ba$_3$NiSb$_2$O$_9$ (6H-B) has been proposed as a QSL candidate\cite{Cheng11}. This system has magnetic ions Ni$^{2+}$ forming triangular layers with spin-1 local moments. The Curie-Weiss temperature is $-75.5 $K and no sign of magnetic ordering is detected down to $0.35 $K, indicating
a strong frustration. The system exhibits the QSL phenomenology described above, with a linear-$T$ specific heat and constant spin susceptibility at low temperatures\cite{Cheng11}. To account for the experiments, Refs.~\onlinecite{Serbyn11} and~\onlinecite{Xu11} proposed QSLs with fermionic spinons. In contrast to these interesting proposals, in this Letter we argue that the 6H-B data can be understood without invoking QSL physics. We propose that the putative QSL behavior arises as a crossover tied to the proximity of a quantum critical point (QCP) between spin spirals favored by the frustrated exchange, and a quantum paramagnetic (QP) phase, favored by a single-ion anisotropy (SIA).  

More specifically, in a mean-field treatment we find the dispersion of spin excitations has the schematic form $\epsilon_{\bs{k}} = \sqrt{f_1(\bs{k}) f_2(\bs{k}) }$. While generically, including at the QCP, there is no special relationship between the functions $f_1$ and $f_2$, in a broad parameter regime near the QCP $f_1$ and $f_2$ are approximately proportional. This leads to an enhanced DOS, and, due to the presence of a degenerate contour of low-energy excitations, a linear intermediate-temperature specific heat.  This behavior follows from the form of the dispersion, and is expected to be robust beyond mean-field theory (MFT). The microscopic ingredients for this behavior are SIA combined with comparable Ising and transverse antiferromagnetic exchange.  Therefore, we expect that such a deviation from a generic dispersion, accompanied by anomalous intermediate-temperature thermodynamics,  should be common in $S > 1/2$ antiferromagnets where the crystal structure admits a SIA.  More broadly, there are certainly many mechanisms by which generic behavior may be pushed down to very low temperatures, and the resulting regimes of anomalous intermediate-temperature behavior may be important in various situations, perhaps even in other QSL candidates.

\begin{figure}[htp]
\includegraphics[width=7cm]{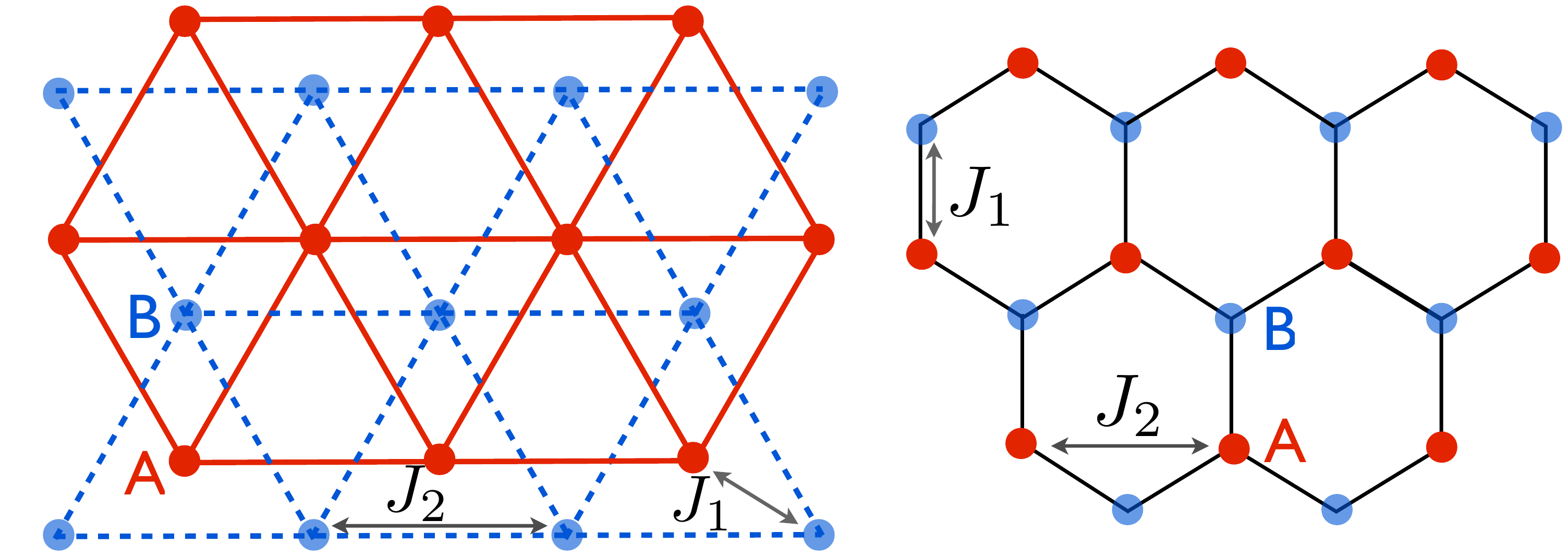}
\caption{(Color online) 
Two adjacent triangular layers of 6H-B (left) can be viewed as a single-layer honeycomb lattice (right). 
$J_1$ ($J_2$) is the interlayer (intralayer) exchange\cite{supplement}.  
$J_1$ is also the exchange between adjacent honeycomb layers.}
\label{fig:lattice}
\end{figure}

In 6H-B, the Ni triangular layers have an A-B stacking with the lattice sites on one layer projecting to the centers of the triangle plaquettes on 
the two neighboring layers (Fig.~\ref{fig:lattice}). Our minimal model includes the interlayer and intralayer spin exchange and 
a SIA. Treating the two neighboring triangular layers as the two sublattices of a honeycomb lattice, we view the 
system as a multilayer honeycomb lattice (Fig.~\ref{fig:lattice}).
 Therefore, when the exchange is dominant and frustrated, the classical ground state is highly degenerate. 
Quantum fluctuations lift the degeneracy and favor coplanar spiral orders. A strong easy-plane SIA favors a QP 
state, which is separated from the ordered state by a QCP. We propose that 6H-B is close to this QCP, and may lie either on the QP or magnetically ordered side. The 
constant spin susceptibility arises from the explicit breaking of spin rotational symmetry by the SIA,  and the powder nature of the samples. 
More notably, we interpret the observed broad linear-$T$ specific heat in terms of the dispersion for spin excitations, as discussed above.

\emph{Model}---Although the interlayer exchange 
path goes through one more oxygen than the intralayer coupling, the multiplicity of the former path is larger than the latter. Moreover, in a structurally 
similar material 6H-A-Ba$_3$NiSb$_2$O$_9$ with long-range magnetic order, the magnetic specific heat at low temperatures is observed to behave as $C_v(T) \sim T^3$, 
which indicates a non-negligible interlayer coupling. That may thus be important in understanding the properties of 6H-B, but is not required for our theory. The resulting exchange is 
given on the triangular multilayers by the Hamiltonian, 
\begin{equation}
{\mathcal H}_{\text{ex}}  =
J_1 \sum_{\langle ij \rangle \in \text{AB}} {\bf S}_i \cdot {\bf S}_j + J_2 \sum_{\langle ij \rangle \in \text{AA and BB} } {\bf S}_i \cdot {\bf S}_j .
\end{equation}
The first sum is for interlayer exchange between nearest-neighbor (NN) sites on neighboring A and B layers, and the second sum is for intralayer
exchange between NN sites within the same layer.  As illustrated in Fig.~\ref{fig:lattice}, the interlayer (intralayer) exchange on a triangular bilayer can be viewed as the 
nearest-neighbor (next-nearest-neighbor) exchange on a single honeycomb layer. 
In contrast to Ref.~\onlinecite{Serbyn11}, we do not include the biquadratic exchange, 
which we expect to be strongly subdominant to ${\mathcal H}_{\text{ex}}$.

The space-group symmetry P6$_3$mc of 6H-B restricts the SIA to be 
${\mathcal H}_{\text{ani}} = D \sum_i (S_i^z )^2$ with $z$-axis normal to the triangular layers. Since an easy-axis anisotropy is more likely to favor magnetic order, 
so we expect easy-plane anisotropy ($D > 0$) for 6H-B, where such order is not observed.

Our model thus contains two competing terms, exchange and SIA, 
${\mathcal H}  =  {\mathcal H}_{\text{ex}}  +  {\mathcal H}_{\text{ani}}.$ Implementing high-$T$  series expansion, we extract the Curie-Weiss temperature, finding that 
$\Theta_{\text{CW}}^z =- D/3- 4J $ and $\Theta_{\text{CW}}^{\perp} =D/6-4J $ (where $J \equiv J_1 + J_2$) for field applied along and 
perpendicular to the $z$ axis, respectively. With a powder sample in experiment\cite{Cheng11}, a powder average gives 
$\Theta_{\text{CW}}^{\text{av}} = - 4 J$ that is independent of $D$. 
Furthermore, 
with Weiss-MFT we demonstrate that saturation temperature of spin susceptibility (observed to be $\sim 25$K\cite{Cheng11}) is set by $D$, that is thus comparable to $J$. 

For the Hamiltonian ${\mathcal H}$, when the SIA dominates with $D \gg J$, 
the ground state is a uniform  QP state with $|S^z = 0 \rangle$ at each site. In the opposite limit of dominant 
exchange, we expect the ground state to be magnetically ordered. Luttinger-Tisza method\cite{Luttinger46} gives the classical ground state 
spin configurations with the ordering wavevector $q_z =0$ and spins lying in the $xy$ plane. When $J_1 > 3J_2$, the classical ground state 
is a usual N\'{e}el state. When $J_1 < 3 J_2$, the classical ground state is degenerate with degenerate spiral wavevectors ${\bf q}_{\perp} 
\equiv (q_x, q_y)$ satisfying
$\sum_{  \{ {\bf b}   \}} \cos ({\bf q}_{\perp} \cdot {\bf b} ) = (\frac{J_1}{J_2} )^2-3$,
where $\{{\bf b}\}$ are 6 next-nearest-neighbor lattice vectors of the honeycomb lattice. The degenerate wavevectors form contour curves 
in momentum space. Moreover, with vanishing $J_1$, this spiral reduces to the familiar 120$^o$ state
of decoupled triangular layers. Quantum fluctuations lift the degeneracy of these classical spin spirals,  selecting states 
characterized by a discrete set of ${\bf q}$'s around which the quantum zero-point energy is minimized. The spiral ground states favored 
by the quantum fluctuations do not vary upon introducing the SIA. The optimal spiral wavevectors are given by\cite{Mulder10}
\begin{eqnarray}
{\bf q}_{\perp} &=& \Big(0, \frac{2}{\sqrt{3}}\cos^{-1} \big( (\frac{J_1}{2J_2})^2 -\frac{5}{4} \big)  \Big),  \,\, \text{for}\,  1<\frac{J_1}{J_2}<3
\\
{\bf q}_{\perp} &=& \big(2\cos^{-1} (\frac{J_1}{2J_2}+\frac{1}{2}) , \frac{2\pi}{\sqrt{3}} \big), \quad\quad \text{for}\, \frac{J_1}{J_2}<1,
\end{eqnarray}
and their symmetry equivalents. 

\begin{figure}[htp]
\centering
\includegraphics[width=5.8cm]{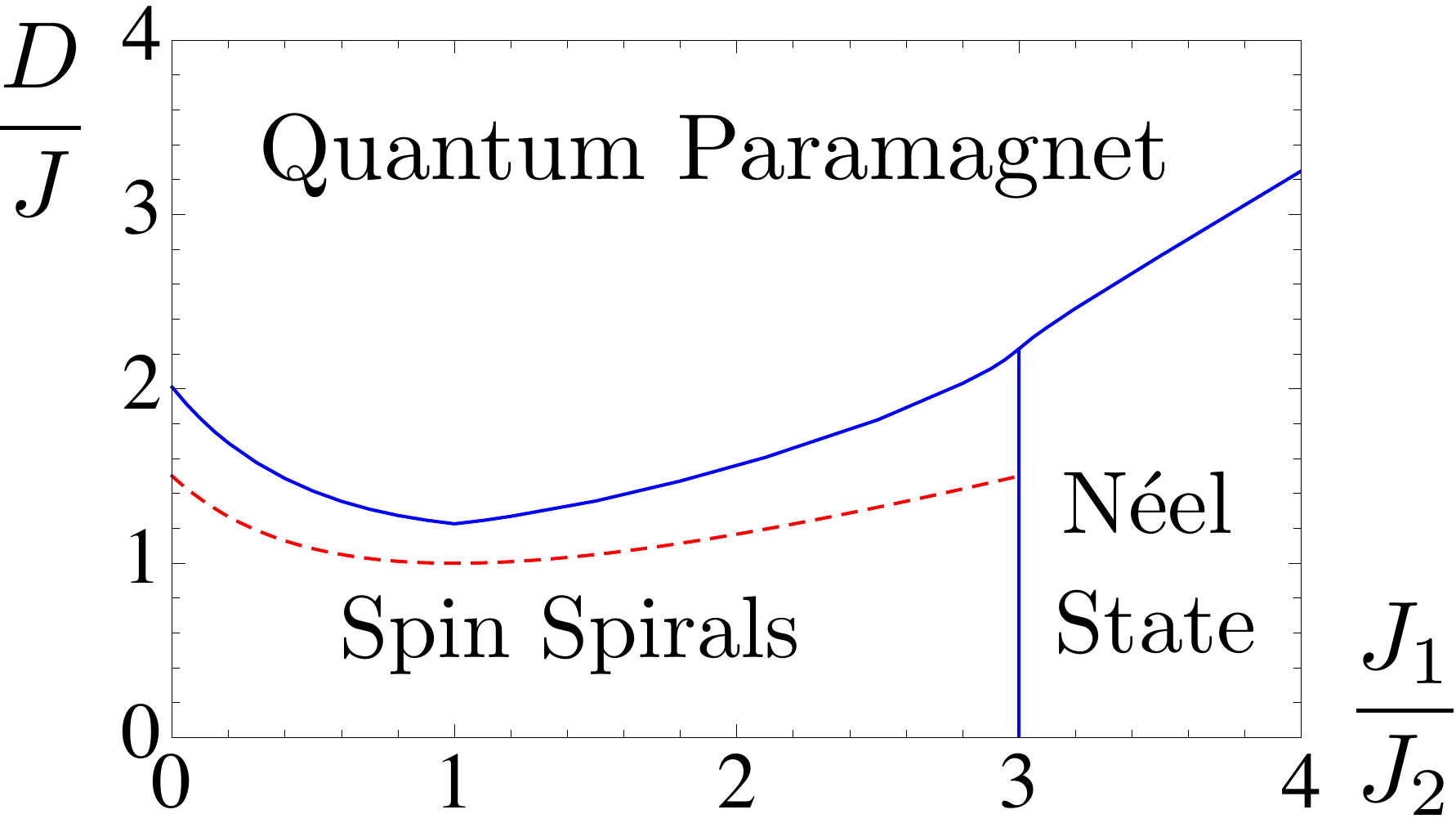}
\caption{Zero-temperature phase diagram determined from the SPE.
Dashed curve indicates the location where $D = \Delta_0/2$, which is important in the discussion of $T$-linear $C_v(T)$ below.}
\label{fig:pd}
\end{figure}

Starting from the magnetically ordered phase, the existence and properties of the phase 
transition can be analyzed within a Weiss-MFT. We decouple the exchange 
into an effective Zeeman field which is then self-consistently determined for each sublattice. We parameterize the spin order as,
\begin{eqnarray}
{\bf S}_{\text A} ({\bf r}) & = & M [\cos ( {\bf q}\cdot {\bf r} )   \hat{x}   + \sin ({\bf q} \cdot {\bf r}) \hat{y} ],
 \\
{\bf S}_{\text B} ({\bf r}) & = & M [ \cos ( {\bf q}\cdot {\bf r} + \theta)   \hat{x}   + \sin ({\bf q} \cdot {\bf r} + \theta) \hat{y} ],
\end{eqnarray}
where $\theta$ is the relative phase between two sublattices and depends on $\frac{J_1}{J_2}$, and $M$ is the order parameter to be determined. This parameterization describes both the N\'{e}el state for $J_1>3 J_2$
and the spin spirals for $J_1 < 3J_2$, with the 120$^o$ state as the limiting case of the decoupled triangular layers.
At zero temperature MFT yields that in the vicinity of the QCP the order parameter is 
$M = \sqrt{2} (1- \frac{D}{D_c} )^{\frac{1}{2}}$ with the critical anisotropy parameter $D_c = 12 (J_1 - J_2)$ for the N\'{e}el state when $J_1>3J_2$, $D_c = 6 J_2$ for the 120$^o$ state at 
vanishing $J_1$, and $D_c = 6J_2 + \frac{2J_1^2}{J_2}$ for the spin spirals when $J_1 <3 J_2$.  We expect that as usual Weiss-MFT overestimates $D_c$ (Fig.~\ref{fig:pd}).

Within Weiss-MFT, in the QP phase, the zero-temperature spin susceptibility
 $\chi^z=0$ (fields along $z$-axis). For fields in $xy$ plane, the spin susceptibility saturates to a constant
$\chi^{ \perp}_0 = \frac{2\mu_0 (g\mu_B)^2}{D+ 12 J}$. The powder average gives the zero-temperature susceptibility $\chi_0^{\text{av}} = 2 \chi_0^{ \perp} /3 $.

\emph{Rotor MFT}---It is convenient to model this easy-plane system with rotor variables, by introducing an integer-valued field $n_i$ and $2\pi$-periodic phase variable $\phi_i$, 
which satisfy $[\phi_i, n_j ] = i\delta_{ij}$. With the mapping ($S^z_i \rightarrow n_i, S^{+}_i \rightarrow \sqrt{2} e^{i \phi_i}$), 
the rotor Hamiltonian reads
\begin{equation}
{\mathcal H}_{\text{rotor}} =  \sum_{ij}  {J_{ij} } [  \cos (\phi_i - \phi_j) + n_i n_j /2] +\sum_i D n_i^2,
\end{equation}
where $J_{ij}$ takes $J_1$ ($J_2$) for NN interlayer (intralayer) bonds. Although $n_i$ only takes the values of $\pm 1, 0$ in the spin model, due to the substantial anisotropy $D$, we expect that relaxing this restriction is unlikely to have significant effects.

Using the coherent-state path integral, we integrate out the field $n_i$ and obtain the partition function,
\begin{equation}
{\mathcal Z} =\int {\mathcal D} \Phi {\mathcal D} \lambda \, e^{-{\mathcal S} -i \sum_i \int d\tau \lambda_i ( |\Phi_i|^2 -1   )} ,
\end{equation}
where ${\mathcal S} =\int d\tau  \sum_{\bf k}  ( 4D \mathbb{I} + 2{\mathcal J}_{\bf k}  )^{-1}_{\mu\nu}   \partial_{\tau} \Phi_{\mu,{\bf k}}^{\ast} \partial_{\tau} \Phi_{\nu, -{\bf k}} 
+\sum_{ij} J_{ij} \Phi_i^{\ast} \Phi_j$ with 
$\Phi_i^{\ast} \equiv  e^{i \phi_i}$. ${\mathcal J}_{\bf k}$ is the $2 \times 2$ exchange coupling matrix in momentum space, $\mu,\nu$ are the sublattice indices, and $\mathbb{I}$ is a $2\times 2$ identity matrix. The constraint $|\Phi_i|=1$ is enforced by the Lagrange multiplier $\lambda_i$.  We proceed by a saddle-point approximation. Assuming $i \lambda_i = \beta \Delta (T)$ at the saddle point, we integrate
out the $\Phi$ field and obtain the saddle-point equation (SPE) for $\Delta (T)$ in paramagnetic phase,
\begin{equation}
\sum_{i=\pm} \int_{ {\bf k} \in \text{BZ}} \frac{ d^3 {\bf k} }{u_{\text{BZ}}} \frac{2D + s_{i,{\bf k}} }{ \epsilon_{i,{\bf k}} } \coth (\frac{\beta \epsilon_{i,{\bf k}}}{2} ) =2,
\label{eq:saddle}
\end{equation}
where $u_{\text{BZ}} =\frac{16\pi^3}{\sqrt{3}}$ is the Brillouin zone (BZ) volume, $s_{\pm,\bf k} \equiv  J_2 \sum_{ \{ {\bf b} \} } \cos ( {\bf k}\cdot {\bf b} ) \pm 2|J_1 \cos(\frac{k_z}{2})| 
\sqrt{3+  \sum_{ \{ {\bf b} \} } \cos ( {\bf k}\cdot {\bf b} ) }  $ 
are the eigenvalues of ${\mathcal J}_{\bf k}$, and $\epsilon_{\pm,\bf k}$ are the two spin excitations,
\begin{eqnarray}
\epsilon_{\pm,{\bf k}} & = & \sqrt{ (4D  + 2s_{\pm,\bf k} ) (\Delta (T)  + s_{\pm,\bf k} )  } 
\nonumber \\
&=& \sqrt{ 2 \big[  (s_{\pm,\bf k}  + D + \frac{\Delta(T)}{2})^2 - (D - \frac{\Delta(T)}{2})^2  \big]}.
\label{eq:lowE}
\end{eqnarray}

\begin{figure}[htp]
\centering
\includegraphics[width=8cm]{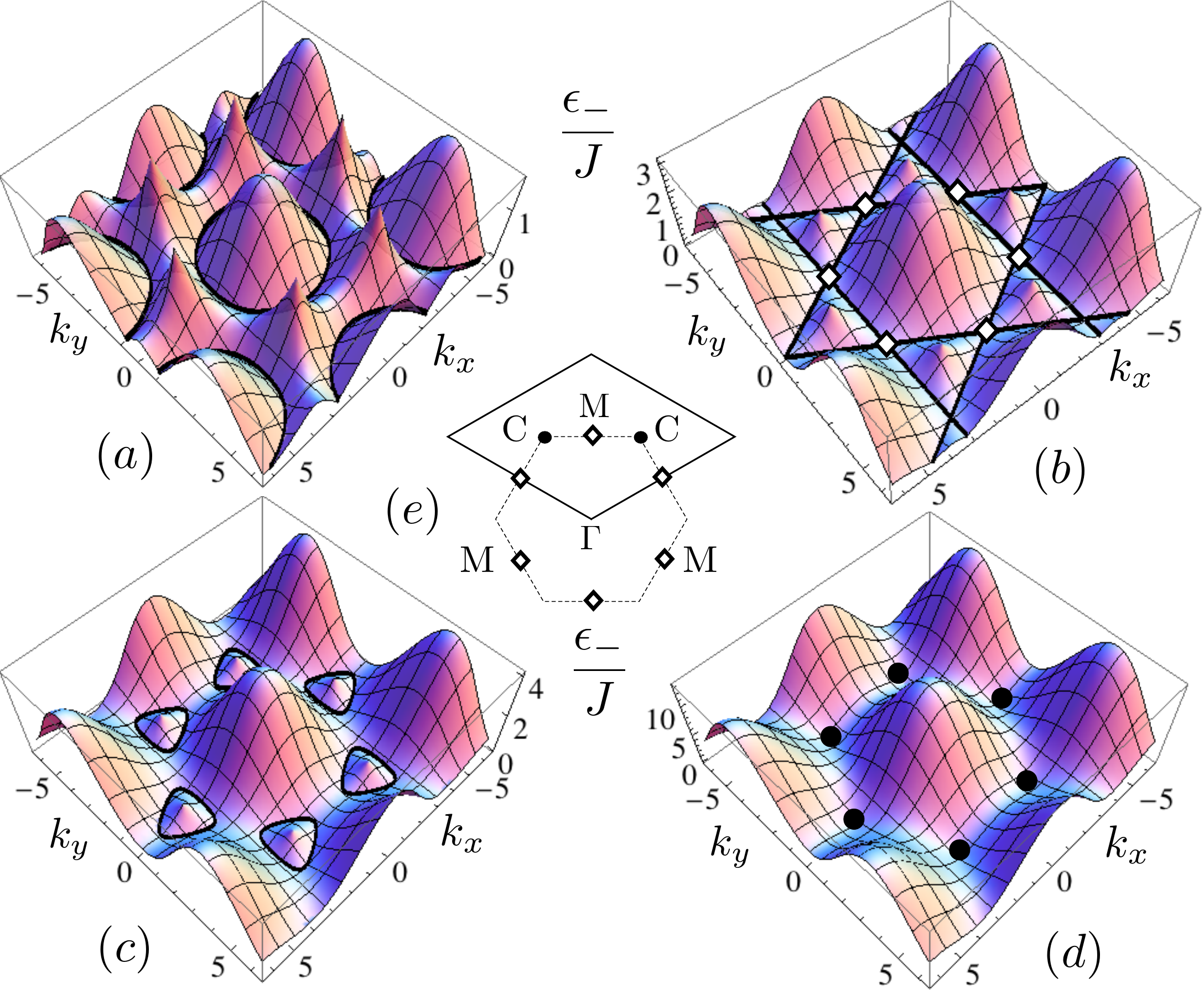}
\caption{(Color online) The evolution of the low-energy excitations in $k_x$-$k_y$ plane with $k_z =0$ at the QCP.  The parameters used in the figures are 
($a$) $J_1= 1.5 J_2, D_c = 1.36  J$,
($b$) $J_1=  J_2, D_c = 1.23 J$,
($c$) $J_1= 0.8 J_2, D_c = 1.28 J$,
($d$) $J_1 =0, D_c = 2.01  J$.
The low-energy gapless contours are marked with bold black lines in ($a$-$c$), while in ($d$) the low-energy gapless points are 
marked with black dots.  Lattice constants are set to 1. ($e$) is the BZ of a honeycomb lattice.
For $J_1 > J_2$, the contour line is centered in the middle of BZ. For $J_1 < J_2$, the contour lines 
are centered around and eventually shrink to the corners of BZ in the limit $J_1 \rightarrow 0$.
The ``$\Diamond$'' in ($b$) correspond to M in ($e$).  }
\label{fig:dispersion}
\end{figure}

When the left-hand side of the SPE is less than $2$ for any $\Delta(T)$, the rotor is condensed which signals magnetic ordering. 
Therefore, besides the transition temperature from the high-temperature paramagnetic phase to the low-temperature spin spirals,  
we also obtain the critical $D_c$ that separates spin spirals from QP phase and the zero-temperature phase diagram(Fig.~\ref{fig:pd}). 
As expected, $D_c$ obtained here is smaller than the one
determined previously from the Weiss-MFT. In particular, $D_c/J$ is minimal at $J_1 = J_2$ corresponding to the largest frustration at this point.
Right at the QCP and zero temperature, $\Delta (0) \equiv \Delta_0 = 3J_2 + J_1^2/J_2$ and the low-energy  mode $\epsilon_{-,{\bf k}}$ develops 
gapless excitations. As shown in Fig.~\ref{fig:dispersion}, the momenta of the gapless excitations form contour lines that are identical to the ones 
of degenerate classical ground state spiral wavevectors. Moreover, as $J_1/J_2$ increases from 0, the contour lines around the BZ corners gradually
expand and meet at M when $J_1 = J_2$.

\begin{figure}[htp]
\includegraphics[width=8.5cm]{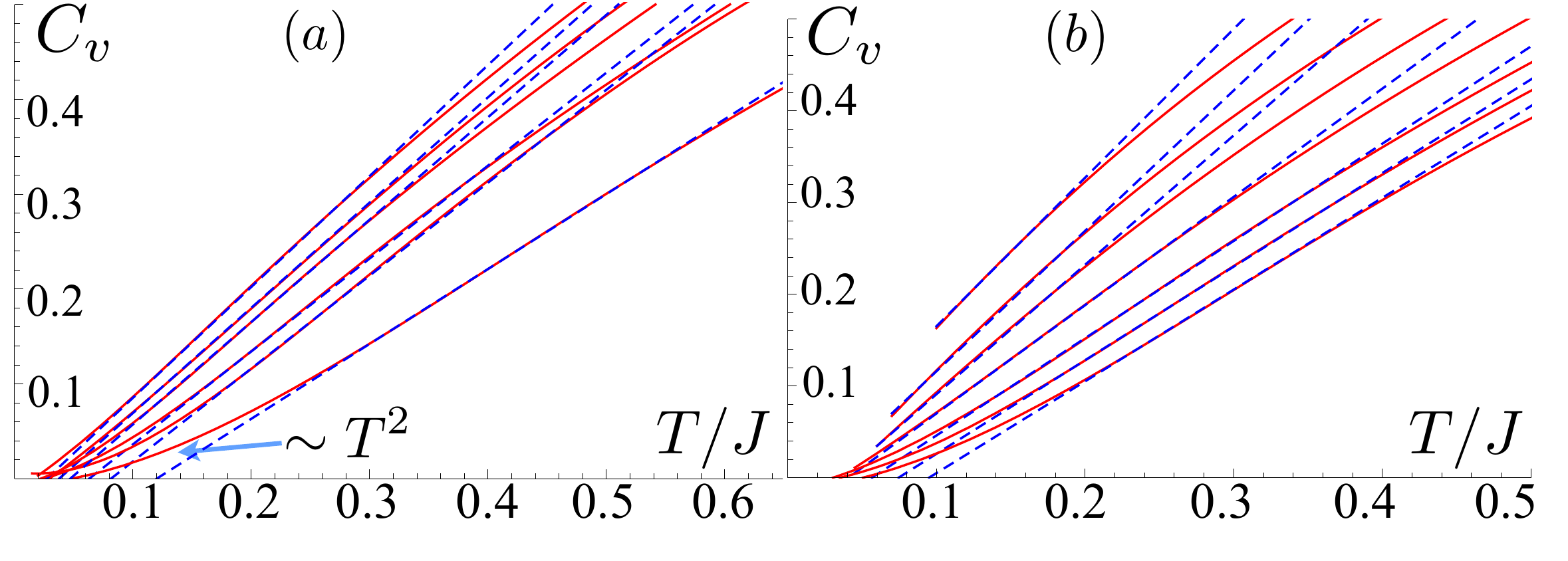}
\caption{(Color online) 
The $C_v$-$T$ plots in the paramagnetic phase. 
In ($a$), from top to bottom, $J_1 =J_2,0.7 J_2,1.5J_2,1.8J_2, 0.3 J_2, 0$ with $D \approx D_c$ and $D=1.24J,1.31J, 1.36J, 1.47J, 1.58J, 2.01J$, respectively. 
In ($b$), $J_1= 0.5 J_2$, $D = 1.08J,1.17J,1.23J, $ $1.32J, 1.41J, 1.48J, 1.55J$ from top to bottom.  For $D=1.08J,$ $1.17J, 1.23J, 1.32J$, $T_c = 0.10J, 0.07J, 0.064J, 0.04J$, respectively.
For $J_1= 0.5 J_2$, $D_c =1.41J$. 
The dashed lines are the linear fits for a range of data points. Energy is in units of $J$.} 
\label{fig:cv}
\end{figure}

\emph{Near the QCP with $T\ll J$}---$\Delta(T)$ increases with $T$ and we define $\Delta(T) \equiv \Delta_0 + \Delta_1(T)$.
The excitation $\epsilon_{\pm}({\bf k})$ picks up a self-energy via the $T$-dependence of $\Delta (T)$.  
By numerically solving the SPE, we find that, near the QCP $\Delta_1 (T) \propto T^2$ for $T \ll J$.  This is also supported by an analytical argument (Supplemental Material\cite{supplement}), and holds in the quasi-2D limit $J_1 \ll J_2$.  This immediately leads to the internal energy $E \propto T^3$ and hence $C_v \propto T^2$, for $T \ll \Omega$ with $\Omega$ an energy cutoff.
This low-temperature $T^2$-$C_v$ regime is confirmed numerically in Fig.~\ref{fig:cv}($a$). 

In Fig.~\ref{fig:cv}($a$), we also find that, as $J_1/J_2$ moves to the point $J_1 = J_2$ from {\sl either} side, the temperature range of the $T^2$-$C_v$
regime diminishes.  We attribute this to the observation that the zero-temperature DOS at the QCP increases with energy, then saturates to a roughly constant value.  This saturation energy (in units of $J$) is found to be lowest when $J_1 = J_2$.

\emph{Linear-$T$ $C_v$ at intermediate $T$}---In Fig.~\ref{fig:cv}, we find an intermediate-temperature regime
with $C_v \approx c_1 T + c_0$. To explain this, we first note that when $D \approx \frac{\Delta(T)}{2}$, the low-energy excitation is approximately the square root of a perfect square:
\begin{eqnarray}
\epsilon_{-,{\bf k}}  &\approx& \sqrt{2} |s_{-,{\bf k}} + D + {\Delta(T)}/{2}|
\nonumber \\
&\approx& \sqrt{2} ( D+ \frac{\Delta_1(T)- \Delta_0}{2} ) + \frac{k^2_{\perp}}{2m_{\perp,k_0}} + \frac{k_z^2}{2m_{z,k_0}}.
\label{eq:eqd}
\end{eqnarray}
If $D =\frac{\Delta_0}{2} $, such dispersion on contour lines in 3D gives a constant DOS at low energies and, because $\Delta(T)$ is only weakly $T$-dependent in this case, this leads to $C_v \propto T$ at low $T$.  This conclusion also holds in the 2D limit $J_1 \ll J_2$. 
As shown in Fig.~\ref{fig:pd}, $D_c$ is slightly greater than $\frac{\Delta_0}{2}$ and the system at $D =\frac{\Delta_0}{2}$ is ordered at very low $T$. 
Once the system enters into the paramagnetic phase, a linear-$T$ $C_v$ is obtained (see the top curve in Fig.~\ref{fig:cv}($b$)). 
Moreover, this linear-$T$ $C_v$ regime persists even when $D$ is increased to or slightly beyond $D_c$.

\emph{Discussion}---The spin susceptibility is observed to saturate to a constant below $25$K\cite{Cheng11}, which is consistent with our prediction. 
Experiments also find $C_v(T) \simeq \gamma T^{\eta}$ with $\eta \approx 1.0(1)$ for $0.35\text{K}<T<7\text{K}$. Determination of both behaviors relies on subtracting a magnetic impurity contribution, which has a significant effect below about $25$K and 1K for $\chi$ and $C_v$, respectively.  The subtraction procedure for $C_v$ relies on fitting $C_v(T, B=0 {\text T}) - C_v(T, B=9 {\text T} )$ to a Schottky form appropriate for a magnetic impurity contribution.  This procedure may be unreliable, because  $9 {\text T}$ is a large field scale for 6H-B, corresponding in temperature units to $\sim 20 {\text K}$, and the fitted difference of $C_v$ is thus expected to include a significant contribution from bulk Ni moments.  While this complicates interpretation of the $C_v(T)$ for $T \lesssim 1 \text{ K}$, we note that $C_v(T)$ is certainly linear for $1\text{K}<T<7\text{K}$ (or $0.05J<T<0.34J$).

As illustrated in Fig.~\ref{fig:cv} for some parameter values, the range of the linear-$T$ specific heat of the experiments is compatible with our results for a range of $J_1/J_2$\cite{Cheng11}.
Taking $J_1 =0.5  J_2$ and $D=1.32 J $ (Fig.~\ref{fig:cv}(b)), we find the powder-averaged zero-temperature spin susceptibility $\chi_0^{\text{av}} \approx 0.0124$emu/mol, which is very close
to the experimental value $0.013$emu/mol. Moreover, the cofficient $\gamma$ in $C_v(T)$ is found to be $204$mJ/mol-K$^2$
 while the experimental value is $168$mJ/mol-K$^2$. We obtain a Wilson ratio of $4.4$, not far from the experimental value $5.6$.
The agreement can be further improved by adjusting $J_1/J_2$ and $D$. 

6H-B may lie either on the QP or magnetically ordered side of the QCP. At very low temperatures we thus expect either a small energy gap, or the onset of spin order. It should be noted that the presence of magnetic impurities may interfere with observation of such very-low-$T$ behavior.  To detect the energy gap or spiral order, NMR or $\mu$SR measurements may be helpful. If 6H-B is in the QP phase, then in a single crystal sample we predict $\chi^z = 0, \chi^{\perp} = \text{const}$ at zero temperature. Ref.~\onlinecite{Serbyn11} considered a state with gapped $S^z = \pm 1$ fermions and gapless $S^z = 0$ fermions forming a Fermi surface. This state has a spin gap, with thermodynamic properties dominated by the Fermi surface. The two QSLs in Ref.~\onlinecite{Xu11} have gapless spin excitations. These states are thus distinct from the QP phase of our proposal, which has a fully gapped spectrum.  
Especially if single crystal samples are available, inelastic neutron scattering should be able to further distinguish these proposals by measuring the dispersion of low-energy spin excitations; our prediction is depicted in Fig.~\ref{fig:dispersion}.

To summarize, we propose a minimal $J_1$-$J_2$-$D$ model for 6H-B and argue that its putative QSL phenomenology is due 
to the proximity to a QCP. Our theoretical prediction is broadly compatible with current experiments\cite{Cheng11}.  
Various future experimental directions are suggested.

\emph{Acknowledgement}--- We thank L. Balicas, P. Lee and L. Thompson for discussion. This research is supported by David and Lucile Packard Fundation (G.C. and M.H.) and NSF through grant no. DMR-1001240 (G.C. and L.R.).

\appendix

\begin{widetext} 

\bigskip\bigskip

\begin{center}
{\bf SUPPLEMENTARY MATERIAL}
\end{center}

\begin{figure}[htp]
\includegraphics[width=10cm]{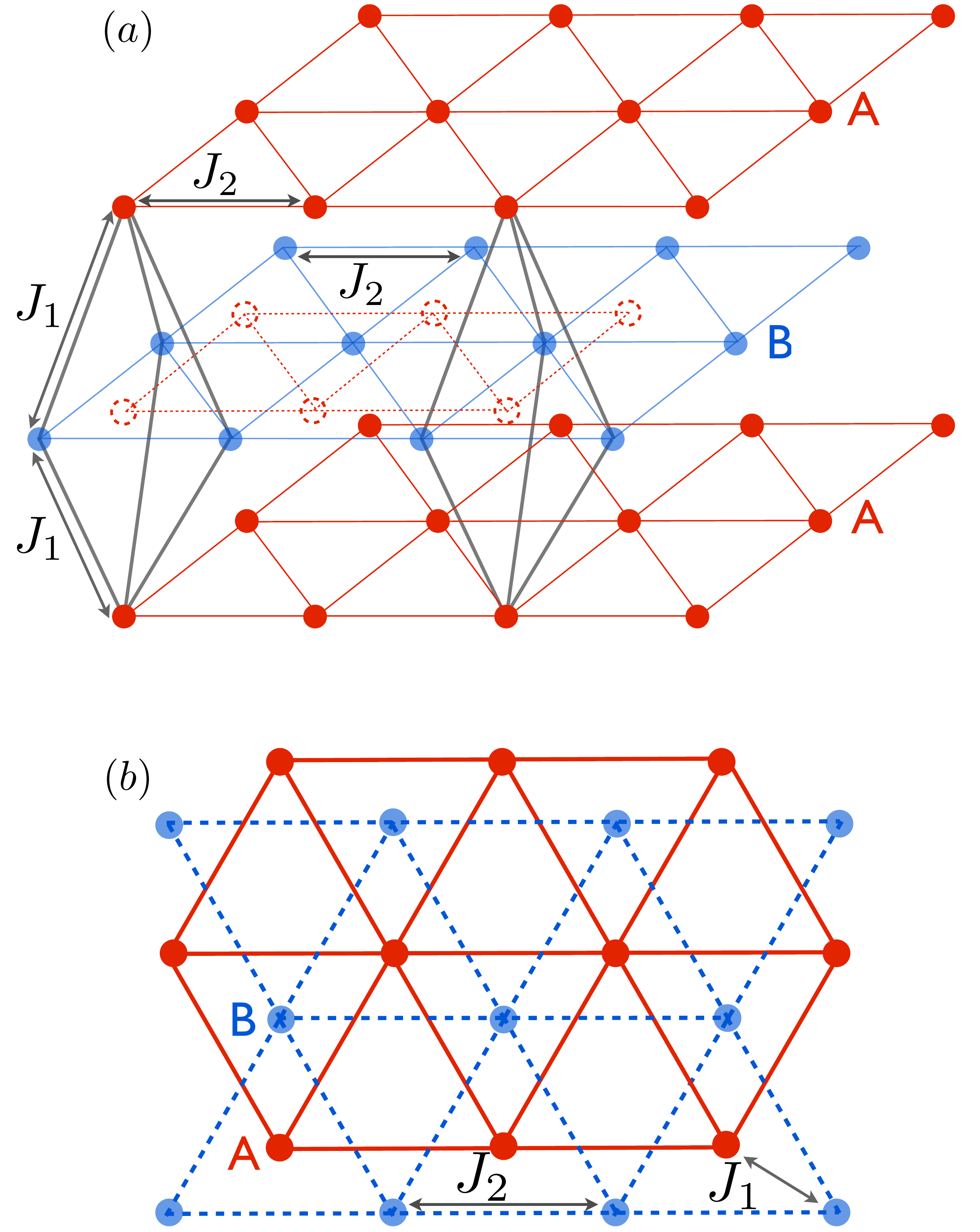}
\caption{(Color online) 
The couplings in the multilayer triangular structure of 6H-B-Ba$_3$NiSb$_2$O$_9$. 
$J_1, J_2$ are the interlayer and intralayer couplings, respectively. 
(a) is the three dimensional view of the lattice structure. The dashed circles on the B plane are the projected positions of the lattice sites from the A plane. (b) is the top view of the lattice
(aslo shown in the main text).  }
\label{fig:mhoneycomb}
\end{figure}

Here we provide a detailed analytical argument for the $T^2$ specific heat at low temperature near the QCP.
At finite temperature, $\Delta(T)$ increases with temperature and we define $\Delta(T) \equiv \Delta_0 + \Delta_1(T)$.
The spin excitation $\epsilon_{\pm}({\bf k})$ also picks up a self-energy via the temperature dependence of $\Delta (T)$.  
By numerically solving the saddle-point equation, we find that, near the QCP $\Delta_1 (T) \propto T^2$ at $T \ll J$.
At $T \ll J$, the low-energy spin excitation near the contour lines can be approximated
as,
\begin{equation}
\epsilon_{-,{\bf k}} \approx \sqrt{ A \Delta_1 (T) + v_{\perp,k_0}^2 k_{\perp}^2 + v_{z,k_0}^2 k_z^2},
\label{eq:eqa}
\end{equation}
where $A = 4D_c - 2\Delta_0$, $k_0$ is a momentum coordinate running along the contour lines, $k_{\perp}$ is normal to the tangent of the contour line at $k_0$, 
and we have neglected the weak temperature dependence of the speeds $v_{\perp}^2, v_z^2$. Eq.~\eqref{eq:eqa} is expected to be a good 
approximation for $\epsilon_- ({\bf k})$ less than a cutoff energy $\Omega$ with $T \ll \Omega \ll J$. The saddle-point equation can be approximated as 
 \begin{equation}
\int_{ k_0, k_{\perp},k_z }^{\Lambda}  \frac{A \coth (\frac{\beta}{2} \sqrt{  A \Delta_1+ v^2_{\perp} k_{\perp}^2 + v_z^2 k_z^2 }) } 
{ 2 \sqrt{  A \Delta_1+v^2_{\perp} k_{\perp}^2 + v_z^2 k_z^2 } } + b  =2,
\label{eq:eqb}
\end{equation}
where the integral is over the region around the contour lines with $|k_{\perp}|, |k_z| \lesssim \Lambda$, and $b$ is the approximately $T$-independent contribution from outside this region. At low temperature, the temperature-dependent part of the integral becomes independent of the cutoff $\Omega$, and only depends on $T$ via the dimensionless parameter $\frac{A \Delta_1(T)}{T^2}$.
In order for the integral to be constant in temperature, we thus expect $ \Delta_1(T) \propto T^2$ in the limit $T \ll \Omega$. This result immediately leads to the internal energy, which can be approximated
as 
\begin{equation}
E \sim \int_{ k_0, k_{\perp},k_z }^{\Lambda}  \frac{\sqrt{ A \Delta_1 (T) + v^2_{\perp} k_{\perp}^2 + v_z^2 k_z^2}}
{ e^{\beta \sqrt{ A \Delta_1 (T) + v^2_{\perp} k_{\perp}^2 + v_z^2 k_z^2}} -1  } \propto T^3
\end{equation}
for $T \ll \Omega$. This gives $C_v \propto T^2$ in this temperature regime.

\end{widetext}

\end{document}